\documentclass[pre]{revtex4}
\usepackage{setspace}
\usepackage{graphicx}
\usepackage{dcolumn}
\usepackage{amsmath}
\usepackage{amsfonts}
\usepackage{amssymb}
\usepackage{subfigure}
\begin{document}

\title{Parallel versus off-pathway Michaelis-Menten mechanism for single-enzyme kinetics of a fluctuating enzyme}
\author{Ashutosh Kumar, Hiranmay Maity and Arti Dua}
\affiliation{Department of Chemistry, Indian Institute of Technology, Madras, Chennai-600036, India}

\date{\today}

\begin{abstract}
\noindent
Recent fluorescence  spectroscopy measurements of the turnover time distribution of single-enzyme turnover kinetics of $\beta$-galactosidase provide evidence of Michaelis-Menten kinetics at low substrate concentration. However,  at high substrate concentrations,  the dimensionless variance of the turnover time distribution shows systematic deviations from the Michaelis-Menten prediction. This difference is attributed to conformational fluctuations in both the enzyme and the enzyme-substrate complex and to the possibility of both parallel and off-pathway kinetics.  Here, we use the chemical master equation to model the kinetics of a single fluctuating enzyme that can yield a product through either  parallel or off-pathway mechanisms. An exact expression is obtained for the turnover time distribution from which the mean turnover time and randomness parameters are calculated.  The parallel and off-pathway mechanisms yield strikingly different dependences of the  mean turnover time and the randomness parameter on the substrate concentration. In the parallel mechanism, the distinct contributions of enzyme and enzyme-substrate fluctuations are clearly discerned from the variation of the randomness parameter with substrate concentration. From these general results we conclude that an off-pathway mechanism, with substantial enzyme-substrate fluctuations, is needed to rationalize the experimental findings of single-enzyme turnover kinetics of $\beta$-galactosidase.
\end{abstract}
\maketitle

\eject

\section{Introduction}
\noindent

Several biological processes that sustain life depend crucially on the catalytic activity of enzymes. A simple reaction mechanism for enzyme-catalysed  reactions was first proposed in 1913 by  Michaelis and Menten$^{1}$ following the work of Wurtz, Henri and several others.$^2$ According to the Michaelis-Menten (MM) mechanism, enzyme $E$ binds with substrate $S$ to form an enzyme-substrate complex $ES$, which either dissociates irreversibly to form product $P$, regenerating the free enzyme $E$ or dissociate reversibly to release the substrate:
\begin{equation}\label{mmm}
 E + S \mathop{\rightleftharpoons}^{k_1}_{k_{-1}} ES  \xrightarrow{k_2}  E + P 
\end{equation}
In the above mechanism, $k_1$ and $k_{-1}$ are the forward and backward rate constants for reversible substrate binding and substrate release steps respectively and $k_2$ is the rate constant for irreversible product formation and enzyme regeneration step.

The classical MM kinetics, based on deterministic mass action kinetics, estimates the temporal variation of the concentrations of enzyme, enzyme-substrate complex and product once the initial concentrations are known.$^{3,4}$  The deterministic nature of kinetics implicitly assume that all enzymes react at the same time to form products in bulk. In the steady-state, therefore, the initial rate of  product formation is given by the classical Michaelis-Menten (MM) equation,
\begin{equation}\label{mm-eq}
v=\frac{k_2 [E]_{0}[S]}{[S]+K_M},
\end{equation}
where $v$ is the enzyme velocity. In the above expression, $ [E]_{0}=[E]+[ES] $ is the initial enzyme concentration and $ K_M=(k_{-1}+k_2)/k_1 $ is the Michaelis constant. When the scaled reciprocal rate $[E]_{0}/v$ is plotted against the reciprocal concentration $1/[S]$, the MM equation yields a linear curve with the slope and intercept given by $K_M/k_2$ and $1/k_2$ respectively. This transformation, suggested by Lineweaver and Burk$^{5}$ in 1934, is widely used to obtain the kinetic parameters of enzyme-catalyzed biochemical reactions. 

A crucial step in the Michaelis-Menten mechanism is the binding of the enzyme with a substrate to from an enzyme-substrate complex, dissociation of  which leads to the formation of product.  An enzymatic conformational state is, however, not fixed but undergoes incessant fluctuations due to random collisions with the surrounding solvent molecules. The latter impart enough thermal energy to the enzyme to pass through the energetically unfavourable transition states that separate different conformational states.  In an enzyme catalyzed reaction, therefore, different conformational states of an enzyme or an  enzyme-substrate complex represent different kinetic intermediates which can interconvert between each other on time scales slower or faster than the product formation (or enzyme turnover) time.$^{6-9}$  How does the rate of  enzymatic conformational fluctuations influence the turnover time of a single enzyme? Does the MM kinetics of a single fluctuating enzyme shows deviation from the classical MM equation? Do these fluctuations about the mean turnover time represent irrelevant random noise or carry useful information about the mechanistic pathway that can be characterized quantitatively? These are some of the questions that have just begun to be addressed in  several recent studies on single-enzyme catalysis.$^{6-19}$

In a recent single-enzyme turnover experiment  based on fluorescence spectroscopy,$^{8,9}$ the turnover time for the repeated turnover of $\beta$-galactosidase has been measured over a long time interval. From the frequency of occurrence of turnover times, the turnover time distribution of the enzymatic turnovers has been obtained, the first moment of which yields the mean turnover time.  The turnover time distribution is mono-exponential at low substrate concentration and multi-exponential at high substrate concentration. Remarkably, the reciprocal of the mean turnover time exactly recovers the classical MM equation at low substrate concentration, $1/\left< t \right> = v/[E]_0$, implying that the mean turnover time obtained from the single-enzyme turnover time distribution is related to the (ensemble-average) enzymatic velocity in the classical MM kinetics. At high substrate concentration, however,  the turnover time distribution shows multi-exponential decay resulting in systematic deviations from the MM kinetics. In particular, the dimensionless variance, which is expected to be unity for (single-pathway) MM kinetics is found to be greater than unity at high substrate concentration. This difference is attributed to conformational fluctuations in both the enzyme and the enzyme-substrate complex and to the possibility of both parallel and off-pathway MM kinetics.

In the past few years, the randomness parameter has found widespread use as a new kinetic parameter to characterize temporal fluctuations in single-enzyme kinetics.$^6$ The randomness parameter, which is  the dimensionless ratio of the variance and the square of the mean $r = \frac{\left<t^2\right> - \left<t\right>^2}{\left<t\right>^2}$, provides a statistical measure of temporal fluctuations in the underlying mechanism in terms of the number of kinetic intermediates in the reaction pathway and their connectivity with respect to each other.$^{6,18}$ Interestingly, a reaction mechanism with $n$ sequentially connected kinetic states, $1 \xrightarrow{k} 2 \xrightarrow{k} 3 \xrightarrow{k} \cdots \xrightarrow{k} n \xrightarrow{k} n+1 $, where occupancy time for each kinetic state is assumed to be exponentially distributed $\omega(t) = k e^{-k t}$, always yields $r =  1/n$ for going from $1$ to $n+1$ state.$^{20,21}$ Here, $n$ represents the number of rate determining steps in a sequential mechanism with the same rate constant $k$.  Thus, $r=1/n$  is the minimum amount of randomness that can be captured by the randomness parameter in a sequential mechanism.$^6$ This implies that the values of $r  < 1$ or $r = 1$ signify multiple or single rate determining step(s) respectively. The opposite limit of $ r > 1$, on the other hand, implies off-pathway states as in inhibited enzyme mechanism.$^{16}$ The randomness parameter can become greater than unity even when product formation occurs via parallel-pathway mechanism due to enzymatic interconversion. The latter mechanism has been proposed to rationalize the results of a recent single-enzyme turnover experiment on $\beta$-galactosidase, where the randomness parameter increases from unity to greater than unity with the increase in substrate concentration.$^{8,9}$  At high substrate concentration, it saturates to a constant value greater than unity.

The following two-state parallel-pathway MM mechanism  serves as a minimal model to observe the effects of enzymatic conformational fluctuations in the MM kinetics:$^{8,9}$
\begin{eqnarray}\label{pmm}
& &E_1 + S  \mathop{\rightleftharpoons}^{k_{11}}_{k_{-11}} ES_1 \xrightarrow{k_{21}}   E_1^0 + P, E_1^0 \xrightarrow{\delta_{21}} E_1 \nonumber\\
& & {\alpha} {\upharpoonleft \downharpoonright}{\alpha}~~~~~~~~~ {\beta} {\upharpoonleft \downharpoonright}{\beta}~~~~~~ {\gamma} {\upharpoonleft \downharpoonright}{\gamma}\nonumber\\
& & E_2 + S \mathop{\rightleftharpoons}^{k_{12}}_{k_{-12}}  ES_2 \xrightarrow{k_{22}}  E_2^0 +P,  E_2^0 \xrightarrow{\delta_{22}} E_2 
\end{eqnarray} 
In this mechanism, a single enzyme, enzyme-substrate intermediate  and the regenerated enzyme at any time $t$ can exist in any one of the two conformer states given by $E_1$ or $E_2$;  $ES_1$ or $ES_2$ and $E_1^0$ or  $E_2^0$ respectively. Here, $\alpha$, $\beta$ and $\gamma$ are the rate constants specifying interconversion between two enzyme,  enzyme-substrate and regenerated enzyme conformers respectively.

In a recent work,$^{17}$ an approximate expression for the turnover time distribution for the above mechanism has been obtained in terms of the individual turnover time distributions of the first $f_{E_1}(t)$ and second $f_{E_2}(t)$ enzyme conformers,
\begin{equation}{\label{wtd:kou}}
f(t) = w_1 f_{E_1}(t) + w_2 f_{E_2}(t),
\end{equation}
where  $f_{E_1}(t)$ and second $f_{E_2}(t)$ are the turnover time distributions for the independent (single-pathway) MM reaction steps, given by $E_1 \rightleftharpoons ES_1 \rightarrow E_1^0 +P, E_1^0 \rightarrow E_1$ and $E_2 \rightleftharpoons ES_2 \rightarrow E_2^0 + P, E_2^0 \rightarrow E_2$ respectively. 
Also, $w_1$ and $w_2$ are the steady-state probabilities for the enzyme to exist in either $E_1$ or $E_2$ state. This form of expression includes the details of enzyme and enzyme-substrate conformational fluctuations in $w_1$ and $w_2$.  The above distribution shows mono-exponential decay at low substrate concentrations and multi-exponential decay at high substrate concentrations which is in qualitative agreement with the experimental findings. It does not, however, recover the Michaelis-Menten equation [Eq. (\ref{mm-eq})] in the absence of enzymatic conformational fluctuations ($\alpha = \beta  = 0$). In the absence of this limiting behaviour, the conditions under which  enzymatic conformational fluctuations influence the MM kinetics can not be fully characterized. 

In this work, starting from the chemical master equation for a single fluctuating enzyme following the parallel or off-pathway mechanisms, we obtain an {\it exact}  expression for the turnover time distribution, the mean turnover time and the randomness parameter. The mean turnover time and randomness parameter exactly recover the MM kinetics in the absence of enzymatic fluctuations. In the presence of these fluctuations, however, the parallel and off-pathway mechanisms show strikingly different variations of the mean turnover time, and the randomness parameter, with substrate concentration.  Further, the individual contributions of enzyme and enzyme-substrate fluctuations in the parallel-pathway mechanism are clearly discerned from the variation of the randomness parameter with substrate concentration.  From our general analysis, we conclude that off-pathway mechanism with single product formation step in the sole presence of enzyme-substrate fluctuations is required to rationalize the results of the single-enzyme turnover  experiment on $\beta$-galactosidase.

This paper is organized as follows. Section II describes the general features of the parallel-pathway MM mechanism with enzymatic conformational fluctuations. It also outlines the key steps of the calculation. Subsections A, B and C of Section III provide numerical analysis of the analytical expressions in terms of the turnover time distribution, the mean turnover time and  randomness parameter for single-pathway, parallel-pathway and off-pathway MM mechanisms respectively. Section IV presents a brief summary of the results along with conclusions. The calculations are detailed in Appendices A and B.

\section{Single-enzyme MM kinetics in the presence of enzymatic fluctuations}

\noindent

At the single-enzyme level, products are not formed in bulk but one at a time, resulting in a single enzyme turnover in each catalytic cycle. The inherent stochasticity of a chemical reaction ensures that  the repeated turnover of a single enzyme observed over a long time interval yields a  distribution of turnover (or dwell) time.$^{22}$ For the parallel pathway MM mechanism  [Eq. (\ref{pmm})], the intrinsic stochasticity of an enzyme-catalyzed reaction and discrete integer jump in the number of reactants, intermediates and products at time $t$ are accounted for in  the chemical master equation approach (CME)$^{22-26}$  of stochastic processes.The effects of  stochasticity are included by considering the number of enzymes $(n_{E_i})$, enzyme-substrates $(n_{ES_i})$, regenerated enzyme $(n_{E_i^0})$ and products $(n_p)$  as discrete random variables which can only take finite number of  positive integral values as determined by stoichiometry. The CME for the  parallel-pathway MM mechanism  [Eq. (\ref{pmm})] is given by
\begin{eqnarray}\label{cme}
 {\partial_{t} P} &=& 
   \left[ \sum_{i=1}^2 \left( k_{1i}' (\mathbb{E}_{E_i} \mathbb{E}_{E_iS}^{-1} - 1 \right) n_{E_i} +  k_{-1i} (\mathbb{E}_{E_i}^{-1} \mathbb{E}_{E_iS} - 1) n_{E_iS} \right.  \nonumber\\
& & + \left. \left. k_{2i} (\mathbb{E}_{E_iS} \mathbb{E}_{E_p}^{-1} \mathbb{E}_{E_i^0}^{-1} - 1) n_{E_iS}  \right) \right] P +  \alpha \left[ (\mathbb{E}_{E_1} \mathbb{E}_{E_2}^{-1} - 1) n_{E_1} + (\mathbb{E}_{E_1}^{-1} \mathbb{E}_{E_2} - 1) n_{E_2} \right] P \nonumber\\
& & + \beta \left[ (\mathbb{E}_{E_1S} \mathbb{E}_{E_2S}^{-1} - 1) n_{E_1S} + (\mathbb{E}_{E_1S}^{-1} \mathbb{E}_{E_2S} - 1) n_{E_2S} \right] P 
 \end{eqnarray}
 where $P$ represents the joint probability distribution given by $P(n_{E_1}, n_{ES_1}, n_{E_2}, n_{ES_2}, n_{E_1^0}, n_{E_2^0}, n_P; t)$ and $\mathbb{E}$ is the step operator$^{25}$ which operates on an arbitrary function $f(x)$ yielding  $\mathbb{E}  f(x) = f(x+1)$ and $\mathbb{E}^{-1}  f(x) = f(x- 1)$. This implies $(\mathbb{E}_x \mathbb{E}_y^{-1} - 1) x f(x, y, z) = (x+1) f(x+1, y-1, z)- x f(x, y, z)$. In the above equation $k_{11}' = k_{11}[S]$ and $k_{12}' = k_{12}[S]$ are the pseudo first-order rate constants. 
 
The CME is written for the time evolution of the joint probability of the number of each species involved in the chemical reaction.$^{25,26}$ For a single enzyme, at any time $t$, different enzyme (or enzyme-substrate) conformational states are mutually exclusive implying $P_{E_1}(t) = P(1,0,0,0,0,0,0;t)$, $P_{ES_1}(t) = P(0,1,0,0,0,0,0;t)$, $P_{E_2}(t) = P(0,0,1,0,0,0,0;t)$, $P_{ES_2}(t) = P(0,0,0,1,0,0,0;t)$.$^{16}$   As a result, the chemical master equation reduces to the following set of coupled differential equations, given by
\begin{eqnarray}\label{cse}
\frac{dP_{E_1}(t)}{dt}&=&-(k_{11}' + {\alpha})P_{E_1}(t)+k_{-11}P_{ES_1}(t)+{\alpha}P_{E_2}(t) \nonumber \\
\frac{dP_{ES_1}(t)}{dt}&=&k_{11}' P_{E_1}(t)-(k_{-11}+k_{21}+{\beta})P_{ES_1}(t)+{\beta}P_{ES_2}(t) \nonumber\\
\frac{dP_{E_2}(t)}{dt}&=&-(k_{12}'+{\alpha}) P_{E_2}(t)+k_{-12}P_{ES_2}(t)+{\alpha} P_{E_1}(t) \nonumber\\
\frac{dP_{ES_2}(t)}{dt}&=&k_{12}' P_{E_2}(t)+{\beta} P_{ES_1}(t)-(k_{-12}+{\beta}+k_{22}) P_{ES_2}(t) \nonumber\\
\frac{dP_{P}(t)}{dt}&=&k_{21}P_{ES_1}(t)+k_{22} P_{ES_2}.
\end{eqnarray}
As the step $E_{i}^0 \longrightarrow E_{i}$ is considered instantaneous, $P_{E_i^0}\approx 0$ and does not appear in the above equations.$^{16,27,28}$  Given this, the constraint, $P_{E_1}(t) + P_{ES_1}(t) + P_{E_2}(t) + P_{ES_2}(t) = 1$ has to be satisfied at all times. It is to be noted that the above differential equations can also be obtained directly by replacing the ``concentration'' descriptor  in deterministic kinetics with the ``probability'' descriptor. However, in the context of enzymatic fluctuations, different conformational states only have probabilistic interpretation and a rigorous derivation of the above equations is possible only from a CME description of the kinetics.

Since the enzyme turnover event is coupled to the product formation, the turnover time probability of an enzyme turnover to occur between time $t$ and $t+{\Delta}t$ is equal to the time for the product  formation $P$ in the same time interval. This implies $f(t){\Delta}t = {\Delta}P_P(t)$, which  in the limit of ${\Delta}t \rightarrow 0$ yields the following expression for the turnover time distribution:
\begin{equation}\label{wtd}
f(t) = \frac{dP_p(t)}{dt} = k_{21}P_{ES_1}(t) + k_{22}P_{ES_2}(t)
\end{equation}
To obtain turnover time distribution from the above expression, Eqs. (\ref{cse}) can be solved to obtain $P_{ES_1}(t)$ and $P_{ES_2}(t)$. In the next section, we provide a detailed analysis of the turnover time distribution and its statistical moments to understand the role of enzymatic conformational fluctuations in single-enzyme MM kinetics.

\section{Turnover statistics of a single enzyme}

The turnover time or dwell time distribution and its statistical moments such as the mean $\left< t \right>$ and the variance $\left< t^2 \right> - \left< t \right>^2$ carry useful information about the mechanistic pathway of an enzyme catalysed reaction.$^{6,18}$ In subsections (B) and (C) of this section, we analyse the statistical properties of an enzyme catalyzed reaction which follows the parallel-pathway and off-pathway mechanisms respectively. In subsection (A) below, we first recapitulate the results of a single-enzyme catalysed reaction following the single-pathway MM mechanism [Eq. (\ref{mmm})]. 

\subsection{Turnover statistics of a single enzyme following the single-pathway MM mechanism}

The turnover time distribution of a single-enzyme catalysed reaction has been obtained previously from the CME approach,$^{22}$ which yields
\begin{equation}\label{sp-wtd}
f(t) = \frac{k_2 k_1 [S]}{2 A} \left[ e^{(A-B) t} - e^{-(A+B) t} \right],
\end{equation}
where $ A = \sqrt{(k_{1} [S] + k_{-1} + k_2 )^2 - 4 k_1 k_2 [S]}/2$ and $B = (k_1[S] + k_{-1} + k_2)/2$. From the above equation, the first and second moments are given by $\left<t\right> = \int_0^\infty ~dt~ t f(t)$ and $\left<t^2\right> = \int_0^\infty ~dt~ t^2 f(t)$. These expressions yield the mean turnover time and the randomness parameter as$^{17,22}$ 
\begin{equation}\label{mtt-mm}
\left< t \right> = \frac{k_1[S] + k_{-1} + k_2}{k_1 k_2 [S]}
\end{equation}
and 
\begin{equation}\label{r-mm}
r = \frac{\left< t^2 \right> - \left< t \right>^2}{ \left< t \right>^2} = \frac{(k_1[S] + k_{-1})^2 + 2 k_2 k_{-1} + k_2^2}{(k_1[S] + k_{-1} + k_2)^2}
\end{equation}
respectively. Comparison of Eq. (\ref{mtt-mm}) with Eq. (\ref{mm-eq}) yields $1/\left< t \right> = v/[E]_0$ implying that the reciprocal of mean turnover time obtained from the single-enzyme turnover time distribution is related to the (ensemble-average) enzymatic velocity in the classical MM kinetics. The minimum value of $r$ can be obtained using $d{r}/d[S] \left|_{[S]=[S]'} = 0 \right.$ resulting in $[S]' = (k_2 + k_{-1})/k_{1}$. At this substrate concentration, $r_{min} = r([S]')$ yields
\begin{equation}\label{r-min}
r_{min} = 1 - \frac{k_2}{2(k_2 + k_{-1})}.
\end{equation}
In the limiting case of $k_{-1} = 0$, the minimum value of $r$ is given by $r_{min} = r([S]') = 1/2$ with $k_1[S]' = k_2$. This implies that there are two rate determining steps ($n= 1/r_{min} = 2$) corresponding to the substrate binding and the product formation steps respectively.  The above equations will be used to analyse the limiting behaviour of the MM kinetics in the presence of enzymatic fluctuations.

\subsection{Turnover statistics of a single fluctuating enzyme following the parallel-pathway MM mechanism}

\subsubsection{Turnover time distribution}

The turnover time distribution for the parallel-pathway MM [Eq. (\ref{pmm})] mechanism is given by Eq. (\ref{wtd}). To obtain an exact expression for the turnover time distribution,  we first solve the coupled differential equations  [Eq. (\ref{cse})] in the Laplace domain, the details of which are presented in Appendix A. Since the expression for the turnover time distribution in the time domain is unwieldy [Eqs. (\ref{pldt}) and (\ref{pldt1})], we directly evaluate the inverse Laplace transform numerically to obtain $f(t)$. 

Fig. (1) shows the temporal variation of the turnover time distribution for the parameter values reported in the figure caption. The turnover statistics of a single fluctuating enzyme is mainly governed by two distinct time scales - the time scale of interconversion between enzymatic conformational states and the time scale of  product formation which is linked to the enzyme turnover event. In what follows, we refer to enzymatic fluctuations as the ones that originate from interconversion between enzyme [$E_1 \mathop{\rightleftharpoons} E_2$] and enzyme-substrate [$ES_1 \mathop{\rightleftharpoons} ES_2$] conformational states. The enzyme or enzyme-substrate conformational fluctuations respectively refer to the presence of either the former or the latter. 

\begin{figure}[t]
\centering
\includegraphics[trim=6cm 6.0cm 0.2cm 1.0cm,clip=true,scale=0.6]{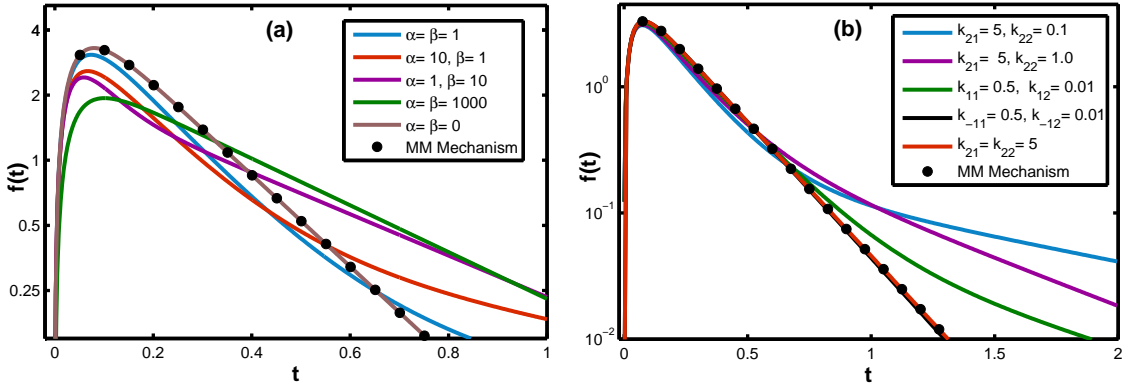}
\caption{Single-enzyme turnover time distribution as a function of turnover time for the parallel-pathway MM mechanism (in non-dimensional units) for (a) different rates of enzyme and enzyme-substrate interconversion characterized by rate constants $\alpha$ and $\beta$ respectively and (b) asymmetry in the rate constants for substrate binding, substrate release and product formation steps.  Common parameter values in (a) and (b) are $k_{11} = k_{-11} = k_{12}^{(b)} = k_{-12}^{(b)} = 0.5$, $k_{21} = 5$, $k_{22}^{(b)} = 0.1$, $\alpha = \beta = 1^{(a)}$ and $[S] = 50$. The parameter values which were varied in subfigures (a) and (b) are indicated as superscripts and the corresponding values are indicated as subfigure legends. Filled circles correspond to the turnover time distribution [Eq. (\ref{sp-wtd})] for the (single-pathway) MM mechanism [Eq. (\ref{mmm})] with $k_{1} = k_{-1} = 0.5$, $k_{2} = 5$ at $[S] = 50$. The latter exactly coincides with the turnover time distribution for the parallel-pathway MM mechanism  (a) in the absence of enzymatic fluctuations $\alpha = \beta = 0$ (brown curve), and (b) symmetry in the catalytic rate constants $k_{21} = k_{22}$ (red curve).}
\end{figure}

Fig. (1a) shows the effect of enzymatic conformational fluctuations on turnover time distribution for parallel-pathway mechanism. In the absence of interconversion between enzymatic conformers, $\alpha = \beta = 0$,  the decay is mono-exponential  [brown curve]. The filled circles represent turnover time distribution [Eq. (\ref{sp-wtd})] for the MM mechanism [Eq. (\ref{mmm})] which exactly coincides with the brown curve indicating that single-pathway MM mechanism is followed in the absence of enzymatic fluctuations. The mono-exponential decay is also observed for the case when enzymatic fluctuations occur faster on the  time scale of the catalytic step [green curve].  Single time scale of decay implies that the effects of fast enzymatic fluctuations are averaged out on the time scale of product formation. 

In between these limits, the decay profiles are multi-exponential implying multiple competing reaction pathways. For the symmetrical case, $\alpha = \beta$, multiexponential decay is observed when enzymatic conformational fluctuations occur slower on the time scale of the catalytic step.  For the asymmetrical case, $\alpha \neq \beta$, multiexponential decay is observed when the time scale of either enzyme or enzyme-substrate interconversion is slower than the catalytic step. All these curves satisfy the condition, $k_{21} \neq k_{22}$. Fig. (1b) exemplifies the latter condition for the symmetrical case, $\alpha = \beta =1$.  When enzymatic conformational fluctuations occur slower on the  time scale of the the catalytic step, then asymmetry in the rate constants for either the catalytic steps $k_{21} \neq k_{22}$  or the substrate binding steps $k_{11} \neq k_{12}$ is necessary to observe multiexponential decay. On the contrary, asymmetry in the substrate release step $k_{-11} \neq k_{-12}$ shows mono-exponential decay [black curve].  The same is also true in the absence of asymmetry in the catalytic steps, which results in mono-exponential decay [red curve] implying that single-pathway MM mechanism is followed [filled circles]. 

Although the present description only accounts for fluctuations between two (instead of $n$) enzyme or enzyme-substrate conformers, it is sufficient to capture the conditions under which the presence of enzymatic fluctuations can result in multi-exponential decay. Fig. (2), for instance, captures the salient features of the turnover time distribution as a function of substrate concentration. This include mono-exponential decay at low substrate concentration and multi-exponential decay at high substrate concentration, in qualitative agreement with the single-enzyme turnover experiment on $\beta$-galactosidase. At low substrate concentration, however, the turnover time distribution for the parallel-pathway MM mechanism [Eq. (\ref{wtd})] does not coincide with the turnover time distribution for the (single-pathway) MM mechanism [Eq. (\ref{sp-wtd})]. This indicates that the MM behaviour is not recovered at low substrate concentration in spite of the mono-exponential nature of the decay curve. The consequences of the latter are discussed below.

It is to be noted that in the context of two-state model of enzymatic conformational fluctuations presented here, the asymmetry in the rate constants for the catalytic  (or substrate binding) step amounts to dynamic disorder in the n-state model implying a broad distribution of $k_2$ (or $k_1$).$^{17}$  

\begin{figure}[t]
\centering
\includegraphics[trim=6cm 6.5cm 0.5cm 1.0cm,clip=true,scale=0.7]{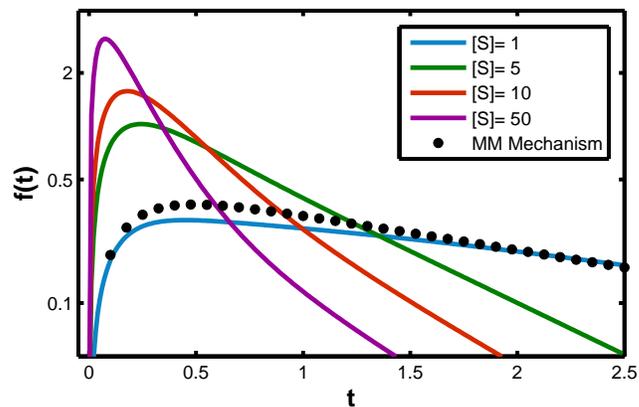}
\caption{Temporal variation of the turnover time distribution for different substrate concentration for the parallel-pathway MM mechanism (in non-dimensional units). The turnover time distribution shows monexponential decay at low substrate concentration and multiexponential decay at high concentration. The parameter values are $k_{11} = k_{-11} = k_{12} = k_{-12} = 0.5$, $k_{21} = 5$, $k_{22} = 1$, and $\alpha = \beta = 1$. Filled circles correspond to the turnover time distribution [Eq. (\ref{sp-wtd})] for the (single-pathway) MM mechanism [Eq. (\ref{mmm})] with $k_{1} = k_{-1} = 0.5$, $k_{2} = 5$ at $[S] = 1$. The latter shows deviation from the turnover time distribution for the parallel-pathway MM mechanism at the same substrate concentration (blue curve).}
\end{figure}

\subsubsection{ First statistical moment and mean turnover time}

We can obtain more statistical information from $f(t)$ by taking its first moment which yields the mean turnover time, ${\langle}t{\rangle} = \int_{0}^{\infty} t f(t) dt$. In the Laplace domain, it can be obtained from ${\langle}t{\rangle} = -\left.\frac{dF(s)}{ds}\right|_{s = 0}$, where $F(s)$ is the Laplace transform of $f(t)$, given by $F(s) = k_{21}P_{ES_1}(s) + k_{22}P_{ES_2}(s)$.  This yields
\begin{eqnarray}\label{mtt}
{\langle} t {\rangle} &=& \frac{{\lambda_3} - k_{21}A_1 - k_{22}A_2}{{\lambda_4}}\nonumber\\
&=& \frac{E[S]^2 + F[S] + G}{H[S]^2 + I[S]}
\end{eqnarray}
where,
\begin{eqnarray}
E &=& k_{11}k_{12}(k_{22}+2{\beta})\nonumber\\
F &=& k_{11}({\alpha}k_{-12}+{\alpha}k_{22}+2{\alpha}{\beta}+k_{-12}{\beta})\nonumber\\
& & + k_{12}(k_{-11}k_{22}+{\beta}k_{-11}+k_{21}k_{22}+k_{21}{\beta}\nonumber\\
& & + k_{22}{\beta}+{\alpha}k_{-11}+{\alpha}k_{21}+2{\alpha}{\beta})\nonumber\\
G &=& 2{\alpha}(k_{-11}k_{-12}+k_{-11}k_{22}+{\beta}k_{-11}+k_{-12}k_{21}\nonumber\\
& & + k_{21}k_{22}+{\beta}k_{21}+{\beta}k_{-12}+{\beta}k_{22})\nonumber\\
H &=& k_{11}k_{12}(k_{21}k_{22}+k_{21}{\beta}+k_{22}{\beta})\nonumber\\
I &=& {\alpha}k_{12}(k_{-11}k_{22}+k_{21}k_{22}+{\beta}k_{21}+{\beta}k_{22})\nonumber\\
& & + {\alpha}k_{11}(k_{21}k_{-12}+k_{21}k_{22}+{\beta}k_{21}+{\beta}k_{22}).\nonumber
\end{eqnarray} 
The enzymatic velocity of single fluctuating enzyme can be obtained from the reciprocal of the mean turnover time,
\begin{equation}
v =  \frac{1}{\langle t \rangle} =\frac{H[S]^2+I[S]}{E[S]^2+F[S]+G}. 
\end{equation}
From the above expression it  can be easily seen that the Michelis-Menten equation, which is characteristically hyperbolic in the substrate concentration [S], does not hold in the presence of enzymatic conformational fluctuations. However, there are certain limiting conditions under which the above equation recovers the Michaelis-Menten equation or the Michaelis-Menten like equation : \\
(i) ${\alpha} = {\beta} = 0$ exactly recovers the Michaelis-Menten equation, $v = \frac{H' [S]}{E' [S] + F'}$, \\
where $H' = k_{11} k_{21}$, $E' = k_{11}$ and $F' = k_{-11} + k_{21}$. \\
(ii) ${\alpha} = 0$  yields the Michaelis-Menten like equation given by $v = \frac{H [S]}{E [S] + F'}$, \\
where $F' = \beta [ k_{11} k_{-12} + k_{12} (k_{-11} + k_{21} + k_{22}) ] + k_{12} (k_{-11} k_{22} + k_{21} k_{22})$\\
(iii) ${\alpha} = 0$ and $\beta \rightarrow \infty$ yields the Michaelis-Menten like equation given by $v = \frac{H' [S]}{E' [S] + F'}$, \\
where $E' = 2 k_{11} k_{12}$, $F' = k_{11} k_{-12} + k_{12} (k_{-11} + k_{21} + k_{22}) $ and $H' = k_{11} k_{12} (k_{21} + k_{22})$\\
(iv) ${\alpha \rightarrow \infty}$  yields the Michaelis-Menten like equation given by $v = \frac{I' [S]}{F' [S] + G'}$,\\
where $F' = k_{11} ( k_{-12} + k_{22} + 2 \beta) + k_{12} ( k_{-11} + k_{21} + 2 \beta)$, $G' = G/\alpha$ and $I' = I/\alpha$.\\
(v) ${\alpha \rightarrow \infty}$ and $\beta \rightarrow \infty$ yields the Michaelis-Menten like equation given by $v = \frac{I' [S]}{F' [S] + G'}$,\\
where $F' =  2 ( k_{11} + k_{12} )$, $G' = 2 ( k_{-11} + k_{-12} + k_{21} + k_{22} )$ and $I' = (k_{21} + k_{22}) (k_{12} + k_{11})$.\\

\begin{figure}[t]
\centering
\includegraphics[trim=4cm 0.1cm 0.2cm 0.1cm,clip=true,scale=0.6]{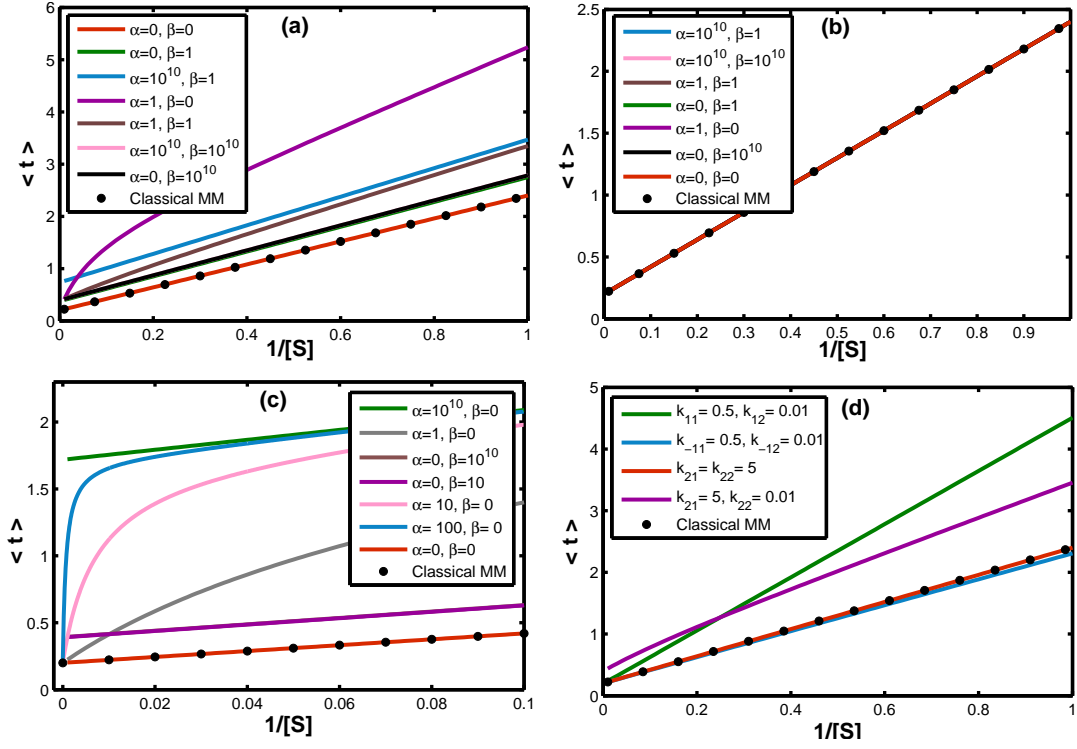}
\caption{Single-enzyme Lineweaver-Burk plots for $\left< t \right>$ versus $1/[S]$ for the parallel pathway MM mechanism (in non-dimensional units) for (a) different rates of enzymatic interconversion characterized by the rate constants $\alpha$ and $\beta$  for $k_{21} \neq k_{22}$, (b) same as (a) but in the absence of asymmetry in the rate constants for the catalytic steps $k_{21} = k_{22}$, (c) different rates of conformational fluctuations arising solely from enzyme or enzyme-substrate interconversion characterized by the rate constants $\alpha$ or $\beta$ respectively and (d) asymmetry in the rate constants for substrate binding, substrate release or catalytic steps. Common parameter values in (a)-(d) are $k_{11} = k_{-11} = k_{12}^{(d)} = k_{-12}^{(d)} = 0.5$, $k_{21} = 5$, $k_{22}^{(b),(d)} = 0.1$ and $\alpha = \beta = 1^{(a),(b),(c)}$. The parameter values which were varied in subfigures (a)-(d) are indicated as superscripts and the corresponding values are indicated as subfigure legends. For the chosen parameter values in (a), the green, black and pink curves almost merge onto each other. Filled circles correspond to the mean turnover time [Eq. (\ref{mtt-mm})] for the (single-pathway) MM mechanism [Eq. (\ref{mmm})] with $k_{1} = k_{-1} = 0.5$, $k_{2} = 5$. }
%    \label{fig:sample_subfigures}
\end{figure}

The above analysis shows that the MM equation is exactly recovered only in the absence of enzymatic conformational fluctuations [condition (i)]. In order to analyse Eq. (\ref{mtt}) in greater detail, we numerically evaluate the mean turnover time $\left< t \right>$ as a function of the reciprocal of the substrate concentration $1/[S]$, which yields the single-enzyme Lineweaver-Burk (LB) curves [Fig. (3)]. 

Fig. (3a) graphically depicts the conditions under which the MM or MM like behaviour can be obtained. In the absence of enzymatic conformational fluctuations, ${\alpha} = {\beta} = 0$ [condition (i)], the MM equation is exactly recovered yielding a linear curve. This is shown as the collapse of the linear red curve with filled circles, obtained from the variation of $[E]_0/v$ with respect to $1/[S]$ in the MM equation [Eq. (\ref{mm-eq})]. The conditions (ii)-(v), represented by green, black, blue and pink curves respectively yield linear curves with intercepts and slopes different from that of the MM equation [red curve], thereby showing the MM-like behaviour. For the symmetrical case, $\alpha = \beta$,  the non-linearity in the LB curve occurs when enzymatic fluctuations occur on time scale longer than the catalytic step [brown curve]. All these curves satisfy the condition, $k_{21} \neq k_{22}$.  When $k_{21} = k_{22}$, all linear and non-linear curves merge with the red curve to exactly recover the MM equation implying that single-pathway MM mechanism is followed under these conditions [Fig. (3b)]. 

Fig. (3c) shows the effect of enzyme or enzyme substrate conformational fluctuations on the mean turnover time [conditions (ii)-(iv)]. In the limiting cases of $\alpha = 0$, $\beta \rightarrow \infty$ and $\alpha \rightarrow  \infty$,  the MM-like behaviour is recovered corresponding to the conditions (iii) and (iv) respectively. The sole presence of enzyme-substrate conformational fluctuations [merged brown and purple curves]  also yields MM-like equation [condition (ii)].  The sole presence of enzyme conformational fluctuations, on the other hand, yields non-linearity in the LB curves even when enzyme conformational fluctuations  occur faster on the time scale of the product formation step [blue, pink and gray curves]. Irrespective of the time scale of enzyme conformational fluctuations [blue, pink and gray curves], the parallel-pathway MM mechanism approaches the single-pathway MM mechanism [red curve] at high substrate concentration.  In the sole presence of enzyme-substrate conformational fluctuations, however, the brown and purple curves do not approach the red curve implying that the parallel-pathway MM mechanism is followed at all substrate concentration.  Once again, all these curves satisfy the condition $k_{21} \neq k_{22}$. 

%The decrease in $\alpha$ decreases the  mean turnover time of the reaction shifting the non-linearity towards lower $[S]$. This is because $E_{1} \leftrightharpoons ES_{1}$ interconversion is preferred over $E_1 \leftrightharpoons E_2$ as $alpha$ become progressively lower. With the increase in  $\alpha$, the latter behaviour is approached at higher values of $[S]$. 

To explore the condition of asymmetry in the rate constants further, Fig. (3d) shows the LB curves for the symmetrical case, $\alpha = \beta = 1$, when enzymatic conformational fluctuations occur slower on the time scale of the catalytic  step. For this condition, asymmetry in the rate constants for the catalytic step shows nonlinearity in the LB curve [purple curve]. Asymmetry in the substrate binding step $k_{11} \neq k_{12}$ shows linear  MM-like behaviour with a shift from parallel to single-pathway mechanism at high substrate concentration [green curve].  Asymmetry in the substrate release step $k_{-11} \neq k_{-12}$, however, shows close agreement with the MM equation [blue curve] implying that single-pathway MM mechanism is followed. 

Figs. (3a), (3c) and (3d) show that the mean turnover time captures the substrate concentration dependent switch from parallel to single-pathway MM mechanism in terms of the approach of the non-linear LB curves or linear MM-like curves towards the exact MM equation [Eq. (\ref{mm-eq})] at high substrate concentration. In next section, we show how the randomness parameter corroborate these results.

\subsubsection{ Second statistical moment and randomness parameter}

\noindent

\begin{figure}[t]
\centering
\includegraphics[trim=5cm 0.08cm 0.2cm 0.08cm,clip=true,scale=0.6]{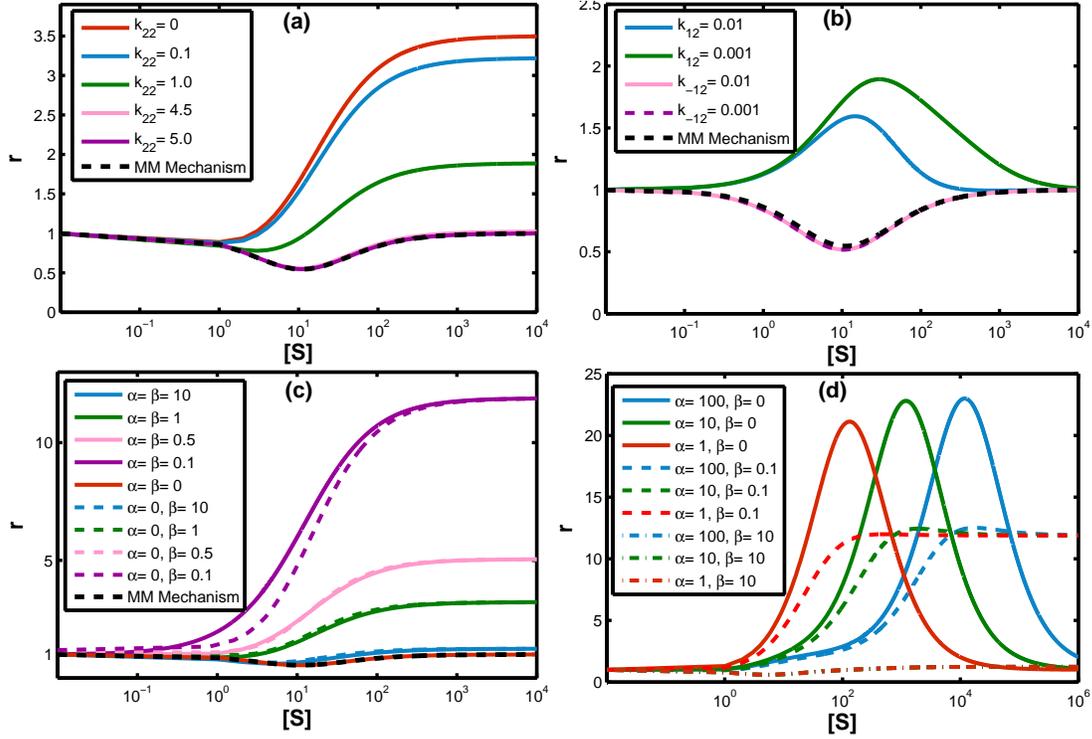}
\caption{Randomness parameter as a function of substrate concentration (in non-dimensional units) for parallel pathway MM mechanism  for (a) asymmetry in the rate constants for product formation step;  (b) for asymmetry in the rate constants for substrate binding and substrate release steps; (c) different enzymatic and enzyme-substrate interconversion rate constants characterized by $\alpha = \beta$ and $\alpha = 0, \beta \neq 0$ respectively and (d) different enzymatic and enzyme interconversion rate constants characterized $\alpha \neq \beta$ and $\alpha \neq 0, \beta = 0$ respectively. Common parameter values in (a)-(d) are $k_{11} = k_{-11} = k_{12}^{(b)} = k_{-12}^{(b)} = 0.5$, $k_{21} = 5$ and $k_{22}^{(a)} = 0.1$ and $\alpha = \beta = 1^{(c),(d)}$. The parameter values which were varied in subfigures (a)-(d) are indicated as superscripts and the corresponding values are indicated as subfigure legends.  Dashed black curves in (a)-(c) correspond to the randomness parameter [Eq. (\ref{r-mm})] for the (single-pathway) MM mechanism [Eq. (\ref{mmm})] with $k_{1} = k_{-1} = 0.5$, $k_{2} = 5$. }
%    \label{fig:sample_subfigures}
\end{figure}

The randomness parameter (or squared coefficient of variation)$^{29,30}$ is a dimensionless ratio of the variance to the mean-square,  $ r = \frac{\langle t^2 \rangle- \langle t \rangle^2}{\langle t \rangle^2} $.  It quantifies the magnitude of temporal fluctuations in a given reaction pathway which can be correlated with the underlying reaction mechanism.   The values of $r = 1$ or $r < 1$ suggest a sequential mechanism with one or more than one rate determining steps respectively. The opposite limit of $r > 1$ can arise if there are off-pathway states with single pathway for product formation or parallel-pathway states with more than one pathway for product formation.

To obtain $r = \frac{\langle t^2 \rangle- \langle t \rangle^2}{\langle t \rangle^2} $, we first evaluate the second moment of $f(t)$, which is given by  ${\langle} t^2 {\rangle} = \int_{0}^{\infty} t^2 f(t) dt$  or ${\langle}t^2{\rangle} = \left.\frac{d^2 F(s)}{ds^2}\right|_{s = 0}$, where $ F(s) = k_{21}P_{ES_1}(s)+k_{22}P_{ES_2}(s)$.  This yields
 \begin{equation}\label{sm}
 {\langle} t^2 {\rangle} = \frac{k_{11}k_{21}-2\lambda_2}{\lambda_4}-\frac{2\lambda_3(k_{21}A_1+k_{22}A_2)}{\lambda_4^2}+\frac{2\lambda_3^2}{\lambda_4^2},
 \end{equation}
where the expressions for $\lambda_2$, $\lambda_3$, $\lambda_4$, $A_1$ and $A_2$ are given in Appendix A. Eqs. (\ref{mtt}) and (\ref{sm}) are then used to obtain $r$. The variation of the randomness parameter with substrate concentration is shown in Figs. (4a)-(4d). A common feature of these figures is that $r = 1$ at low substrate concentration as substrate binding is the rate determining step.
 
Fig. (4a) shows the effect of asymmetry in the rate constants for the catalytic step when enzymatic conformational fluctuations occur slower on the time scale of the catalytic step. For perfectly symmetrical case, $k_{21} = k_{22}$, the substrate concentration dependence of the randomness parameter shows initial decrease from  unity followed by convergence to unity at high substrate concentration. The latter suggests that the single-pathway MM mechanism [dashed black curve obtained from Eq. (\ref{r-mm})] is followed at all substrate concentration [purple curve]. Following the analysis of Eq. (\ref{r-min}), it can be readily seen that the randomness parameter attains the minimum value of $r_{min}  = 0.545$ at $[S]' = 11$ for the parameter values $k_{-11} = k_{-12} = k_{-1} = 0.5$ and $k_{21} = k_{22} = k_2 = 5$. At high substrate concentration, the randomness parameter approaches unity as the product formation step is the rate determining step. For the asymmetrical case, $k_{21} \neq k_{22}$, the randomness parameter first increases from its initial value of unity before saturating to a constant value greater than unity at high substrate concentration. More the asymmetry in the catalytic rate constants, higher the value $r$ at which the saturation occurs. This suggests substrate concentration dependent switch from single-pathway to parallel-pathway MM mechanism for the condition $k_{21} \neq k_{22}$. 

The effect of asymmetry in the rate constants for the substrate binding and the substrate release steps are depicted in Fig. (4b) for the case where enzymatic fluctuations occur slower on the time scale of the catalytic step. Unlike the case of asymmetry in the catalytic rate constants discussed above, asymmetry in the substrate binding rate constants, $k_{11} \neq k_{12}$, shows that the randomness parameter instead of saturating to a constant value greater than unity, decreases to unity at high substrate concentration.  The latter implies that the parallel-pathway MM mechanism switches to single-pathway MM mechanism at high substrate concentration. Asymmetry in the substrate release step $k_{-11} \neq k_{-12}$ [purple curve], almost coincides with the single-pathway MM mechanism [dashed curve] implying that the single-pathway MM mechanism is followed at all substrate concentration. 

Fig. (4c) shows the effect of enzymatic conformational fluctuations for the symmetrical case, $\alpha = \beta$. Single-pathway MM mechanism is followed in the absence of enzymatic fluctuations [$\alpha = \beta = 0$]. The same is also true for the case when the enzymatic conformational fluctuations occur faster on the time scale of the catalytic step [blue curve]. As the time scale of enzymatic conformational fluctuations become slower than the catalytic step, the randomness parameter first increases from its initial value of unity and then saturates to a  constant value greater than unity at high substrate concentration. Slower the time scale of enzymatic fluctuations compared to the catalytic step, more the randomness. These results imply a shift from single-pathway MM mechanism at low substrate concentration to parallel-pathway MM mechanism at high substrate concentration, whenever enzymatic fluctuations [$\alpha = \beta$] occur slower on the time scale of the catalytic step. Interestingly, the sole presence of enzyme-substrate conformational fluctuations ($\alpha = 0$) shows similar trends [dashed curves]. The latter suggest that irrespective of the time scale of enzyme conformational fluctuations, the switch from single to parallel-pathway MM mechanism at high substrate concentration is mainly determined by the competition between the time scales of enzyme-substrate interconversion and product formation steps.

\begin{figure}[t]
\centering
\includegraphics[trim=5cm 0.1cm 0.2cm 0.1cm,clip=true,scale=0.6]{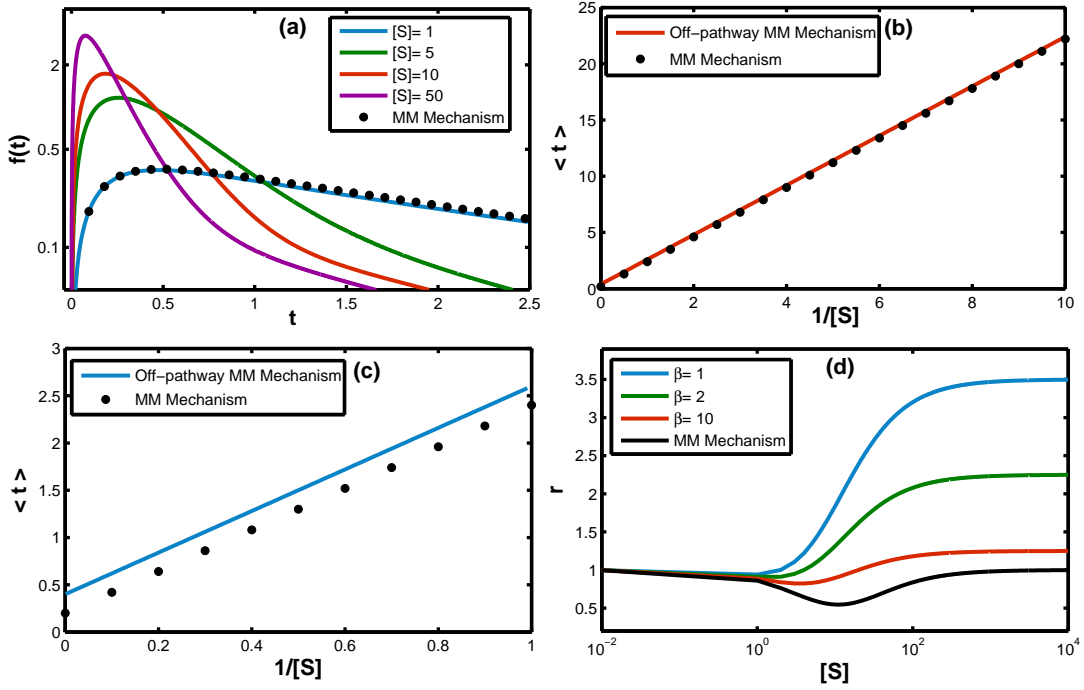}
\caption{Single-enzyme turnover statistics for the off-pathway MM mechanism [Eq. (\ref{mm1})] (in non-dimensional units). (a) Turnover time distribution as a function of time  at different substrate concentrations. (b) Single-enzyme Lineweaver-Burk plot based on the off-pathway MM mechanism [Eq. (\ref{opm-mtt})] shows that the MM equation is recovered at low substrate concentration. (c)  Single-enzyme Lineweaver-Burk plot based on the off-pathway mechanism [Eq. (\ref{opm-mtt})] shows deviation from the MM equation at high substrate concentration. (d) The randomness parameter increases from $r =1$ at low substrate concentration to $r>1$ at high substrate concentration when enzyme-substrate interconversion rate is slower than the catalytic step.  The parameter values for (a)-(d) are $k_{1} = k_{-1}  = 0.5$, $k_{2} = 5$ with $\beta = 1$ in (a). Filled circles in (a)-(c) and black curve in (d) correspond to the turnover time distribution, mean turnover time and randomness parameter for the (single-pathway) MM mechanism obtained from Eqs. (\ref{sp-wtd}), (\ref{mtt-mm}) and (\ref{r-mm}) respectively with parameter values $k_{1} = k_{-1} = 0.5$ and $k_2 = 5$. }
\end{figure}

The asymmetrical case of $\alpha \neq \beta$ shows that the distinct contributions  of enzyme ($\beta=0$) and enzyme-substrate ($\alpha = 0$) conformational fluctuations can be clearly discerned from the variation of the randomness parameter with substrate concentration. Fig. (4d) shows that the sole presence of enzyme conformational fluctuations ($\beta = 0$) results in giant temporal fluctuations, characterized by very high noise-to-signal ratio ($r \gg 1$), even when enzyme conformational fluctuations occur faster on the time scale of the catalytic step. At very high substrate concentration, the randomness parameter becomes unity as the product formation becomes the rate determining step, implying a shift from parallel to single-pathway MM mechanism. Interestingly, even a small increase in the value $\beta$ [dashed curves in Fig. (4d)], suppresses the magnitude of these giant fluctuations to almost half its original value. Also, instead of decaying to unity at very high substrate concentration, the randomness parameter saturates to a constant value greater than unity, implying parallel-pathway  mechanism is followed at high $[S]$.  Thus, in the presence of enzymatic conformational fluctuations with  $\alpha \neq \beta$, the randomness parameter exceeds unity when enzyme-substrate conformational fluctuation occur slower on the time scale of the catalytic step [dashed curves in Fig. (4d)]. In the opposite limit, $r \leq 1$  [dotted curves in Fig. (4d)].

\subsection{Turnover statistics of a single enzyme following the off-pathway MM mechanism}

$\beta$-galactosidase is an enzyme that catalyzes the hydrolysis of lactose and is known to follow the MM kinetics. In a recent experiment based on  fluorescence spectroscopy, the turnover time for the repeated turnover of a single $\beta$-galactosidase was observed over a long time interval from which the turnover time distribution was obtained.$^8$ The latter showed mono-exponential decay at low substrate concentration and multi-exponential decay at high substrate concentration. The single-enzyme Lineweaver Burk plot, obtained from the variation of the mean turnover time with the reciprocal of the substrate concentration,  showed exact agreement with the MM equation [Eq. (\ref{mm-eq})] at low substrate concentration. The variation of the randomness parameter with substrate concentration, on the other hand, was found to be unity at low substrate concentration and greater than unity at high substrate concentration. 

The parallel-pathway MM mechanism, discussed in the previous section, can capture certain features of this experiment. This include (a) mono-exponential decay of the turnover time distribution at low substrate concentration and multi-exponential decay at high substrate concentration [Fig. (2)], and (b) the increase in the randomness parameter from unity to greater than unity with the increase in the substrate concentration [Figs. (4a) and (4c)]. Both these results require that enzymatic [or enzyme-substrate] conformational fluctuations are slower on the time scale of the catalytic step and $k_{21} \neq k_{22}$. However, the same conditions that recover (a) and (b) can not recover the MM equation [Eq. (\ref{mm-eq})] at low substrate concentration. This effect is clearly captured in Fig. (2), which shows that in spite of the mono-exponential decay of the turnover time distribution at low substrate concentration [blue curve], the MM kinetics  [filled circles corresponding to Eq. (\ref{sp-wtd})] is not recovered.

Below, we show that experimental findings can be rationalized on the basis of the following off-pathway MM mechanism:
\begin{eqnarray}\label{mm1}
& &E_1 + S  \mathop{\rightleftharpoons}^{k_{1}}_{k_{-1}} ES_1 \xrightarrow{k_{2}}   E_1^0 + P, E_1^0 \xrightarrow{\delta_{2}} E_1 \nonumber\\
& &~~~~~~~~~~~~~~~ {\beta} {\upharpoonleft \downharpoonright}{\beta}\nonumber\\
& &~~~~~~~~~~~~~~~ES_2 ,
\end{eqnarray} 
where product formation occurs via single pathway mechanism. 

The turnover time distribution for the off-pathway mechanism has been evaluated in Appendix B. Fig. (5a) shows the dependence of the turnover time distribution on substrate concentration. For enzyme-substrate conformational fluctuations occurring slower on the time scale of the catalytic step, the turnover time distribution is mono-exponential at low substrate concentration and multi-exponential at high substrate concentration. However, in contrast to the parallel-pathway mechanism, the turnover time distribution for the off-pathway mechanism at low substrate concentration [blue curve] exactly recovers the MM kinetics [filled circles corresponding to Eq. (\ref{sp-wtd})]. 

The first moment of the distribution yields
\begin{equation}\label{opm-mtt}
v = \frac{1}{\left< t\right>} = \frac{k_1 k_2 [S]}{2 k_1[S] +  (k_{-1} + k_2)},
\end{equation}
which recovers the MM equation at low substrate concentration [Fig. (5b)], but shows deviation from it at high substrate concentration [Fig. (5c)]. With the increase in the substrate concentration, the randomness parameter shows an increase from its initial value of unity to greater than unity whenever enzyme-substrate conformational fluctuations are slower on the time scale of the catalytic step [Fig. (5d)].  The present analysis, thus, shows that an off-pathway MM mechanism, with substantial enzyme-substrate fluctuations, is required to rationalize the results of the single-enzyme turnover experiment on  $\beta$-galactosidase.

\section{Summary and Conclusions}

In this work, we use the chemical master equation to model the kinetics of a single fluctuating enzyme that can yield a product through either parallel or off-pathway mechanisms. We obtain an exact expression for the turnover time distribution, from which the mean turnover time and randomness parameters are evaluated.

For parallel-pathway MM mechanism, the reciprocal of the mean turnover time yields the exact MM equation in the absence of enzymatic conformational fluctuations. 
The presence of enzyme, enzyme-substrate or enzymatic conformational fluctuations, in contrast, yield deviations from the MM equation whenever the rate constants for product formation (or substrate binding) steps are unequal. The variation of the randomness parameter with substrate concentration clearly discerns the distinct contributions of enzyme and enzyme-substrate fluctuations in the parallel-pathway mechanism.  The randomness parameter is found to be greater than unity at high substrate concentration when enzyme-substrate conformational fluctuations are slow on the time scale of the catalytic step, implying that parallel pathway MM kinetics is followed at this concentration. When only enzyme conformational fluctuations are allowed, in contrast, the randomness parameter is greater than unity as long as enzyme conformational fluctuations occur slower on the time scale of the substrate binding step. At high substrate concentration, when the catalytic step becomes the rate determining step, the randomness parameter decays to unity implying the switch from parallel to single-pathway MM kinetics with the increase in the substrate concentration.

The mean turnover time for the off-pathway MM mechanism, allowing only enzyme-substrate conformational fluctuations, yields the exact MM equation at low substrate concentration, but shows deviation from it at high substrate concentration. At these high concentrations, the randomness parameter is greater than unity whenever the enzyme-substrate conformational fluctuations are slow on the time scale of the catalytic step.  

From these general results, we conclude that an off-pathway mechanism with substantial enzyme-substrate conformational fluctuations is needed to rationalize the results of the single-enzyme turnover experiment on  $\beta$-galactosidase. It is conceivable that the turnover statistics of different enzymes reveal different aspects of enzyme and enzyme-substrate conformational fluctuations in single-enzyme kinetics. The present analysis shows that the distinct contributions of enzyme and enzyme-substrate conformational fluctuations in parallel or off-pathway mechanisms can be clearly discerned from the variation of the mean turnover time and the randomness parameter with substrate concentration.
\eject

\appendix
\section{Exact calculation of the turnover time distribution for parallel-pathway MM mechanism}

Eqs. (\ref{cse}) can be solved by taking the Laplace transform. Taking the Laplace transform and applying the initial conditions, $P_{E_1}(0) = 1$, $P_{ES_1}(0) = 0$, $ P_{E_2}(0) = 0$, $P_{ES_2}(0) = 0$ and $P_{P}(0) = 0$,  the coupled differential equations  reduce to the following set of algebraic equations:
\begin{eqnarray}
& & (s+k_{11}' + {\alpha})P_{E_1}(s) - k_{-11}P_{ES_1}(s) -  {\alpha}P_{E_2}(s) = 1\nonumber\\
& & (s+k_{-11} + {\beta}+k_{21})P_{ES_1}(s) - k_{11}'P_{E_1}(s) - {\beta}P_{ES_2} = 0 \nonumber\\
& & (s+k_{12}' + {\alpha})P_{E_2}(s) - k_{-12}P_{ES_2}(s) - {\alpha}P_{E_1}(s) = 0 \nonumber\\
& & (s+{\beta} + k_{-12}+k_{22})P_{ES_2}(s) - k_{12}'P_{E_2}(s) - {\beta}P_{ES_1}(s) = 0
\end{eqnarray}
Rewriting the above equations in the matrix form yields
\[\begin{pmatrix}
(s+k_{11}'+{\alpha}) & -k_{-11} & -{\alpha} & 0 \\ -k_{11}' & (s+k_{-11}+{\beta}+k_{21}) & 0 & -{\beta} \\ -{\alpha} & 0 & (s+k_{12}'+{\alpha}) & -k_{-12} \\ 0 & -{\beta} & -k_{12}' & (s+{\beta}+k_{-12}+k_{22}) 
\end{pmatrix}
\begin{pmatrix}
P_{E_1}(s) \\ P_{ES_1}(s) \\ P_{E_2}(s) \\ P_{ES_2}(s)
\end{pmatrix}=
\begin{pmatrix}
1 \\ 0 \\ 0 \\ 0
\end{pmatrix}
\]
or,\[\begin{pmatrix}
 P_{E_1}(s) \\ P_{ES_1}(s) \\ P_{E_2}(s) \\ P_{ES_2}(s)
\end{pmatrix}=
\begin{pmatrix}
(s+k_{11}'+{\alpha}) & -k_{-11} & -{\alpha} & 0 \\ -k_{11}' & (s+k_{-11}+{\beta}+k_{21}) & 0 & -{\beta} \\ -{\alpha} & 0 & (s+k_{12}'+{\alpha}) & -k_{-12} \\ 0 & -{\beta} & -k_{12}' & (s+{\beta}+k_{-12}+k_{22})
\end{pmatrix}^{-1}
\begin{pmatrix}
1 \\ 0 \\ 0 \\ 0
\end{pmatrix}
\]
 \[\begin{pmatrix}
P_{E_1}(s) \\ P_{ES_1}(s) \\ P_{E_2}(s) \\ P_{ES_2}(s)
\end{pmatrix}=
\begin{pmatrix}
\frac{{(s^3+A_0s^2+B_0s+C_0)}}{(s^4+{\lambda_1}s^3+{\lambda_2}s^2+{\lambda_3}s+{\lambda_4})} \\\frac{{(k_{11}'s^2+A_1s+B_1)}}{(s^4+{\lambda_1}s^3+{\lambda_2}s^2+{\lambda_3}s+{\lambda_4})} \\\frac{{({\alpha}s^2+D_0S+E_0)}}{(s^4+{\lambda_1}s^3+{\lambda_2}s^2+{\lambda_3}s+{\lambda_4})} \\\frac{{(A_2s+B_2)}}{(s^4+{\lambda_1}s^3+{\lambda_2}s^2+{\lambda_3}s+{\lambda_4})}
\end{pmatrix}
\]

where,

\begin{eqnarray}
{A_0}&=&k_{-11}+k_{-12}+k_{21}+k_{22}+2{\beta}+k_{12}[S]+{\alpha}\nonumber\\
{B_0}&=&(k_{-11}+k_{21})(k_{-12}+k_{22})+(k_{-11}+k_{21}+k_{-12}+k_{22}){\beta}\nonumber\\
   & &+(k_{12}[S]+{\alpha})(k_{-12}+k_{22}+2{\beta}+k_{-11}+k_{21})-k_{12}k_{-12}[S]\nonumber\\
{C_0}&=&(k_{12}[S]+{\alpha})[(k_{-11}+k_{21})(k_{-12}+k_{22})+(k_{-11}+k_{21}+k_{-12}+k_{22}){\beta}]\nonumber\\
   &&-k_{12}k_{-12}(k_{-11}+k_{21}+{\beta})[S]\nonumber\\
{D_0}&=&{\alpha}(k_{-11}+k_{-12}+k_{21}+k_{22}+2{\beta})\nonumber\\
{E_0}&=&{\alpha}(k_{-11}+k_{21})(k_{-12}+k_{22})+{\alpha}{\beta}(k_{-12}+k_{22}+k_{-11}+k_{21})+k_{11}k_{-12}{\beta}[S]\nonumber\\
{A_1}&=&k_{11}k_{12}[S]^2+k_{11}(k_{-12}+k_{22}+{\alpha}+{\beta})[S]\nonumber\\
{B_1}&=&k_{11}k_{12}[S]^2(k_{22}+{\beta})+{\alpha}k_{11}[S](k_{-12}+k_{22}+{\beta})+{\alpha}{\beta}k_{12}[S]\nonumber\\
{A_2}&=&(k_{11}{\beta}+k_{12}{\alpha})[S]\nonumber\\
{B_2}&=&k_{11}k_{12}{\beta}[S]^2+k_{12}[S]{\alpha}(k_{-11}+k_{21}+{\beta})+k_{11}{\alpha}{\beta}[S]\nonumber
\end{eqnarray}
and
\begin{eqnarray}
{\lambda_1} &=&(k_{11}+k_{12})[S]+2({\alpha}+{\beta})+k_{-11}+k_{-12}+k_{21}+k_{22}\nonumber\\
{\lambda_2} &=&k_{11}k_{12}[S]^2+k_{11}[S](k_{-12}+k_{21}+k_{k22}+{\alpha}+2{\beta})\nonumber\\
           & &+k_{12}[s](k_{-11}+k_{21}+k_{22}+{\alpha}+2{\beta})\nonumber\\
           & &+(k_{-11}k_{-12}+k_{-11}k_{22}+k_{-11}{\beta}+k{-12}k_{21}+k_{21}k_{22}\nonumber\\
           & &+k_{21}{\beta}+k_{-12}{\beta}+k_{22}{\beta}++2{\alpha}k_{-11}+2{\alpha}k_{21}+2{\alpha}k_{-12}\nonumber\\
           & &+2{\alpha}k_{22}+4{\alpha}{\beta}\nonumber\\
{\lambda_3} &=& k_{11}k_{12}[S]^2(k_{21}+k_{22}+2{\beta})+k_{11}(k_{21}k_{-12}+k_{21}k_{22}+{\beta}k_{21}\nonumber\\
          & &+k_{-12}{\beta}+k_{22}{\beta}+{\alpha}k_{21}+{\alpha}k_{-12}+{\alpha}k_{22}+2{\alpha}{\beta})[S]\nonumber\\
          & &+k_{12}(k_{-11}k_{22}+k_{-11}{\beta}+k_{21}k_{22}+k_{21}{\beta}+k_{22}{\beta}\nonumber\\
          & &+{\alpha}k_{-12}+{\alpha}k_{21}+{\alpha}k_{22}+2{\alpha}{\beta})[S]\nonumber\\
          & &+2({\alpha}k_{-11}k_{-12}+{\alpha}k_{-11}k_{22}+{\alpha}{\beta}k_{-11}+{\alpha}k_{-12}k_{21}\nonumber\\
          & &+{\alpha}k_{21}k_{22}+{\alpha}{\beta}k_{21}+{\alpha}{\beta}k_{-12}+{\alpha}{\beta}k_{22})\nonumber\\ 
{\lambda_4} &=& k_{11}k_{12}(k_{21}k_{22}+k_{21}{\beta}+k_{22}{\beta})[S]^2\nonumber\\
           & &+{\alpha}(k_{-11}k_{12}k_{22}+k_{12}k_{21}k_{22}+k_{12}k_{21}{\beta}+k_{12}k_{22}{\beta}\nonumber\\
           & &+k_{11}k_{21}k_{-12}+k_{11}k_{21}k_{22}+{{\beta}k_{11}k_{21}+{\beta}k_{11}k_{22}})[S]\nonumber            
\end{eqnarray}  
comparing the second and fourth matrix elements of the left hand side with the second and fourth matrix element of the right hand side yields
\begin{eqnarray}\label{pld}
P_{ES_1}(s)&=&\frac{{k_{11}s^2[S]+A_1s+B_1}}{s^4+{\lambda_1}s^3+{\lambda_2}s^2+{\lambda_3}s+{\lambda_4}} \nonumber\\
P_{ES_2}(s)&=&\frac{{A_2s+B_2}}{s^4+{\lambda_1}s^3+{\lambda_2}s^2+{\lambda_3}s+{\lambda_4}}.
\end{eqnarray}
In the Laplace domain, the turnover time distribution is given by
\begin{equation}
f(s)=k_{21}P_{ES_1}(s) + k_{22}P_{ES_2}(s),
\end{equation}
The inverse Laplace transform of Eq. (\ref{pld}) yield,
\begin{eqnarray}\label{pldt}
P_{ES_1}(t)&=& \bigg[\frac{e^{-at}(aA_1-B_1-a^2k_{11}[S])}{(a-b)(a-c)(a-d)}-\frac{e^{-bt}(bA_1-B_1-b^2k_{11}[S])}{(a-b)(b-c)(b-d)}\nonumber \\& & -\frac{e^{-ct}(cA_1-B_1-c^2k_{11}[S])}{(a-c)(c-b)(c-d)} -\frac{e^{-dt}(dA_1-B_1-d^2k_{11}[S])}{(a-d)(d-b)(d-c)} \bigg]
\end{eqnarray}
\begin{eqnarray}\label{pldt1}
P_{ES_2}(t)&=& \bigg[\frac{e^{-at}(aA_2-B_2)}{(a-b)(a-c)(a-d)}-\frac{e^{-bt}(bA_2-B_2)}{(a-b)(b-c)(b-d)}\nonumber \\& & -\frac{e^{-ct}(cA_2-B_2)}{(a-c)(c-b)(c-d)} -\frac{e^{-dt}(dA_2-B_2)}{(a-d)(d-b)(d-c)} \bigg]
\end{eqnarray}
where $a$, $b$, $c$ and $d$ are the effective rate constants, which can be obtained from the numerical solutions of  the quartic equation 
$s^4+{\lambda_1}s^3+{\lambda_2}s^2+{\lambda_3}s+{\lambda_4} = 0$.

\section{Exact calculation of the turnover time distribution for off-pathway MM mechanism}

The turnover time distribution for the off-pathway MM mechanism [Eq. (\ref{mm1})] is given by $f(t) = k_2 P_{ES_1}(t)$. This can be evaluated from Eq. (\ref{cse}) by taking $k_{11}' = k_1'$, $k_{-11} = k_{-1}$, $k_{21} = k_2$ and $k_{12}' = k_{-12} = k_{22} = 0$. The resulting coupled differential equation, when solved in the Laplace domain, yield the following solution for $P_{ES_1}$:
\begin{equation}
P_{ES_1}(s) = \frac{(A_1's+B_1')}{(s^3+{\lambda_1}'s^2+{\lambda_2}'s+{\lambda_3}')}
\end{equation}
where, $k_1' = k_1 [S] $, $A_1'=k_1[S]$, $B_1'= k_1[S]\beta$, ${\lambda_1}'=k_{-1}+k_1[S]+k_2+2{\beta}$, ${\lambda_2}' = k_1[S]k_2+k_{-1}{\beta}+2k_1[S]{\beta}+k_2{\beta}$ and ${\lambda_3}' = k_1[S]k_2{\beta}$. In the Laplace domain, the turnover time distribution and its first and second moments are given by $F(s) = k_2 P_{ES_1}(s)$, $\left< t \right> = -dF(s)/ds|_{s=0}$ and $\left< t^2 \right> = d^2 F(s)/ds^2|_{s=0}$. The last two expressions yield the mean turnover time and randomness parameter [Fig. 5]. In the time domain, on the other hand, the turnover time distribution is given by
\begin{eqnarray}
f(t) &=& k_2 \bigg[\frac{e^{-a't}(-aA_1'+B_1')}{(a'-b')(a'-c')}+\frac{e^{-b't}(A_1'b'-B_1')}{(a'-b')(b'-c')}\nonumber\\
&&+\frac{e^{-c't}(-B_1'+A_1'c')}{(a'-c')(-b'+c')}\bigg],
\end{eqnarray}
where $ a', b' $ and $c'$ are the effective rate constants, which can be obtained from the numerical solutions of the cubic equation $ s^3+{\lambda_1}'s^2+{\lambda_2}'s+{\lambda_3}'=0 $.

\begin{acknowledgments}
AK acknowledges the financial support from the Council of Scientific and Industrial Research (CSIR), Government of India.
\end{acknowledgments}

\end{document}